\documentclass[preprint,amsmath,amssymb,showpacs]{revtex4}

\usepackage{amssymb}
\usepackage{float}
\usepackage{graphicx}
\usepackage{dcolumn}
\usepackage{amsmath}
\usepackage{subfigure}
\usepackage[hypcap]{caption}
\usepackage{hyperref}
\usepackage{epstopdf}
\usepackage{bm}
\usepackage[latin1]{inputenc}
\newcommand{\comment}[1]{}

\begin{document}




\title{Mixing in modulated turbulence. Numerical results}


\author{Yuyao Yang$^1$, Robert Chahine$^1$, Robert Rubinstein$^2$, Wouter J.T. Bos$^1$}

\affiliation{$^1$ LMFA, CNRS, Ecole Centrale de Lyon, Universit\'e de Lyon, 69134 Ecully, France\\
  $^2$ Newport News, VA, USA
}

\begin{abstract}
Direct numerical simulations are carried out to investigate scalar mixing in an isotropic turbulent flow with a time-periodic forcing. For high amplitudes of the modulation, it is shown that the average mixing rate is negatively affected at low frequencies. In this limit the mixing time scale increases, whereas the typical velocity timescale decreases. We further determine the frequency response of scalar statistics to  a periodic scalar-forcing. 
\end{abstract}

\maketitle


\section{Introduction}

The mixing rate of a scalar quantity advected by a fluid is a key quantity in a wide range of applications. Increasing the mixing rate by changing the flow properties can have far reaching consequences in process-optimization. Whereas the mixing in laminar flows can often be studied analytically, and the mixing rate can be greatly enhanced by changing the boundary conditions \cite{Gouillart2007} or the time-dependence of the flow \cite{Aref1984}, the turbulent case is in general far more complicated. If any understanding of the modification of turbulent mixing through time-dependent forcing is to be obtained, we think it is compulsory to look at the most simplified case. We consider therefore the academic case of periodically forced isotropic turbulence, advecting a passive scalar. 

The response of a turbulent velocity field on a time-periodic isotropic forcing has received a considerable interest since the beginning of the 2000s. The initial studies aimed at identifying a possible resonance in the energy transfer process \cite{Lohse2000,Heydt2003-1,Heydt2003-2}. Direct numerical simulations \cite{Kuczaj2006,Kuczaj2008} and experiments \cite{Cadot2003,Cekli2010} were carried out to systematically investigate the response of the kinetic energy and dissipation rate on the forcing frequency. An analytical study, using two-point closure techniques, allowed to explain the different scaling regimes of the time-dependent quantities \cite{Bos2007-3}. The ability of engineering models to reproduce the different features of periodically forced turbulence was investigated in \cite{Rubinstein2009}. 

The application of particular forcing schemes to influence turbulent mixing was considered in \cite{Kuczaj2006-2}. In that study the forcing was introduced in different wavenumber bands in Fourier space to mimic the complex nature of turbulent flows generated by realistic objects. The influence of the so-generated flow on turbulent mixing was assessed by monitoring the wrinkling of level-sets of an advected scalar. Those results inspired several experimental investigations with application to turbulent combustion (see for instance \cite{Verbeek2011,Verbeek2015}).
In \cite{Kuczaj2006-2} the spatial character of the forcing was modified, but no temporal modulation was applied. In the experiments the modulation was both spatial and temporal, and it is not straightforward to disentangle the different effects, so that it is not clear whether the observed effects were caused by the time-periodic nature of the experimental inlet conditions or the spatial complexity of the latter.

In the present study we consider the influence of a temporal modulation of the forcing on the mixing of a passive scalar. Two different cases are investigated. The first case corresponds to a stationary injection of scalar into a periodically forced velocity field. The second one is the case where, in a statistically steady isotropic turbulence, the scalar is introduced by a modulated injection. We will investigate both cases here by means of direct numerical simulation.  In particular, we will show that in the case of a strong velocity modulation not only the periodic part, but also the time-average of the scalar variance and kinetic energy depend on the modulation frequency.

\section{Definitions and equations}

We consider the Navier-Stokes equations for incompressible flow $\bm u$, mixing a passive scalar $\theta$: 
\begin{eqnarray}
		\frac{\partial }{{\partial t}}\bm{u} + \bm{u} \cdot \nabla \bm{u} =  - \nabla P + {\rm{\nu }}{\nabla ^2}\bm{u} + \bm{f}\\
		\nabla  \cdot \bm{u} = 0\\
		\frac{\partial }{{\partial t}}\theta  + \bm{u} \cdot \nabla \theta  = D{\nabla ^2}\theta  + g \label{eq:theta}
\end{eqnarray}
where $P$ is the pressure (normalized by a uniform density), $\nu$ and $D$ are kinematic viscosity and diffusivity, respectively. The flow and the scalar field are  kept in a statistically stationary state through  an energy and scalar variance input $\bm f$, $g$. The forcing is applied to the smallest wavenumbers of the Fourier-transformed velocity and scalar field, denoted by $\hat u_i$ and $\hat\theta$, respectively,
\begin{eqnarray}
\hat f_i(\kappa,t)=\frac{1}{N_f}\frac{\hat u_i(\bm \kappa,t)}{\left|\hat {\bm  u}(\bm \kappa,t) \right|^2}(\overline p+\tilde p\cos(\omega t))&\\
\hat g(\kappa,t)=\frac{1}{N_f}\frac{\hat \theta(\bm \kappa,t)}{\left|\hat \theta(\bm \kappa,t) \right|^2}(\overline p_\theta+\tilde p_\theta\cos(\omega t)),&~~~~~~~~~~\left| \bm{\kappa} \right| \le {\kappa_F},
\end{eqnarray}
with $N_F$ the total number of forced modes. The quantities $p$ and $p_\theta$ denote the kinetic energy and scalar variance injection rates. Overlined quantities denote time-averages and tilded quantities denote periodic fluctuations around the temporal average. In applying such forcing schemes, a statistically steady state is observed where the phase averaged energy and scalar injection fluctuate periodically around the time-averaged values,
\begin{eqnarray}
 \left<\bm f_i \bm u_i \right>=\overline{p} +  \tilde p\cos(\omega t), \\
 \left<g \theta  \right>=\overline{p}_{\theta} + \tilde p_\theta\cos(\omega t).
\end{eqnarray}
These forcing schemes will result in a statistically isotropic velocity and scalar field. In our case the flow domain is a spatially periodic box. In such a setting the evolution equations for the kinetic energy $k=\frac{1}{2}\left< |\bm u|^2\right>$  and the variance of the scalar  $k_\theta=\frac{1}{2}\left< \theta^2\right>$ reduce to
\begin{eqnarray}
 \frac{dk}{dt}=p-\epsilon\\
 \frac{dk_\theta}{dt}=p_\theta-\epsilon_\theta.
\end{eqnarray}
In these equations $\epsilon$ and $\epsilon_\theta$ are the phase averaged dissipation of kinetic energy and scalar variance, respectively. All the different statistical quantities of interest will in the following be decomposed into a time-averaged and a periodic component. The time-averaged balance equations for the kinetic energy and scalar variance are given by
\begin{eqnarray}
0=\overline P-\overline\epsilon\\
0=\overline P_\theta-\overline \epsilon_\theta,
\end{eqnarray}
and the periodic quantities evolve as
\begin{eqnarray}
-\omega \tilde k \sin(\omega t+\phi_k)=\tilde p \cos(\omega t)-\tilde \epsilon \cos(\omega t+\phi_{\epsilon}),\\
-\omega \tilde k_\theta \sin(\omega t+\phi_{k_\theta})=\tilde p_\theta \cos(\omega t)-\tilde \epsilon_\theta \cos(\omega t+\phi_{\epsilon_\theta}). 
\end{eqnarray}
In these expressions, we have assumed that all quantities will periodically oscillate around a mean value with a period $\omega$. In previous works, \cite{Heydt2003-1,Heydt2003-2,Kuczaj2006,Bos2007-3,Kuczaj2008}, the frequency dependence of $\tilde k$ and $\tilde \epsilon$ was investigated. 

We will start by reproducing these results in section \ref{sec:flucU}. After that we will investigate the case where the relative periodic forcing amplitude $\alpha_p=\tilde p/\bar p$ is large enough to not only influence the periodic quantities, but also the time-averaged kinetic energy and scalar variance,  $\overline k$ and $\overline k_\theta$. These results show that by adding a large-amplitude modulation to the forcing, one can directly influence the transfer-rate of the kinetic energy $\chi$, and the mixing rate of the passive scalar $\chi_\theta$. These quantities can be defined as the inverse of the integral timescale and scalar timescale, respectively,
\begin{equation}
 \chi=\frac{\overline \epsilon}{\overline k}\textrm{~~and ~~} \chi_\theta=\frac{\overline \epsilon_\theta}{\overline k_\theta}.\label{eq:mixeff}
\end{equation}
In section \ref{sec:flucT}, we will determine the frequency dependence of the periodically fluctuating scalar quantities, $\tilde k_\theta$, $\tilde \epsilon_\theta$ for the case where the scalar injection rate is modulated.

\section{Simulations and postprocessing}

A standard pseudospectral method is used to compute the velocity and scalar field in a space-periodic cubic domain of size $2\pi$.  A conventional $2/3$ wavenumber truncation is used to eliminate the aliasing error and a third order Runge-Kutta, Total Variation Diminishing scheme is used as time discretisation. The same code was used in ref. \cite{Bos2015-1} to study the mixing of temperature fluctuations in isotropic turbulence. 

Simulations are carried out at two different Reynolds numbers. First, low resolution simulations at a spatial resolution of $64^3$ are performed with kinematic viscosity $\nu=0.009$, corresponding to a Taylor Reynolds number $Re_{\lambda}=32$, with eddy turn-over time $T=1.949$. The resolution allows to resolve the smallest scales upto  $k_{max} \eta=0.93$. Another set of simulations is carried out at a resolution of  $N^3 =256^3$ gridpoints, kinematic viscosity $\nu = 0.0009$; Taylor Reynolds number $Re_{\lambda}=105$; eddy turn-over time $T=2.317$ and $k_{max} \eta=0.97$. In all simulations the Schmidt number $Sc \equiv \nu/D= 1$. The forced modes correspond to $0.5<|\bm \kappa|\le 2.5$.

A challenge in the study of the frequency response of turbulent flows is the convergence of the statistics. At low forcing frequencies the simulations become very long if a sufficient number of periods is to be resolved. At high frequencies the response to a periodic forcing will be shown to be small, so that also in this case long simulations are needed, not to resolve sufficient periods, but to be able to distinguish the frequency response from the turbulent fluctuations. Obtaining converged statistics is therefore challenging in both the small and large frequency limits. 

Previous investigations \cite{Kuczaj2006,Bos2007-3} focused in particular on the linear response of turbulence on a periodic modulation. In this limit linearized equations around a given equilibrium allow to analytically derive certain results. The verification of such results is not straightforward in the nonlinear regime, where the perturbation is large. Ideally, to investigate the linear response of a turbulent flow, the amplitude of the forcing should be chosen small compared to the amplitude of the steady part of the forcing ($\alpha_p\equiv\tilde p/\bar p\ll 1$). However, since the turbulent fluctuations are in this case much larger than the periodic response, very long simulations should be carried out to obtain an estimate of the frequency response. In particular at large frequencies, where the frequency response will be shown to drop rapidly as a function of frequency this would impose prohibitively long computations. A compromise is to consider a larger modulation amplitude. In this study, as in \cite{Kuczaj2006}, we use $\alpha_p\equiv\tilde p/\bar p=0.2$. This allows to obtain converged statistics for a large range of frequencies at a reasonable computational cost for low Reynolds number ($R_\lambda=32$). For higher Reynolds number this already leads to prohibitively long simulations. Therefore we have carried out another set of simulations with a relative modulation amplitude $\alpha_p=1$. Even though this certainly violates the linear perturbation assumption, we will show that the frequency-response of the modulated quantities is not quantitatively altered. We will further show that this has an interesting direct influence on the time-averaged quantities.

Before extracting the frequency response of the simulations, the flow was simulated for approximately 10 eddy turn-over times to obtain a statistically steady state. It proved convenient to determine the amplitude of the periodic response by using a Fourier-transform of the signal. Before Fourier-transforming the time-series of a given quantity, a hanning window function is applied to the signal to eliminate the aliasing error at high frequencies due to the finite length of the signal. In the frequency spectra, if the simulations are carried out for a sufficiently long time-interval, the amplitude of the periodic response is easily identified by a sharp peak. Comparing the value of this peak to the neighbouring values in the spectrum gives a direct estimate of the signal-to-(turbulent)-noise ratio.  In all simulations the value of the peak at the considered frequency was at least ten times the value of the neighbouring values in the spectra. For the phase averaged amplitudes, error-bars are added to the datapoints in the figures, computed from the signal to noise ratio. In most cases, this error-bar is smaller than the size of the symbols used in the figures and is then omitted.  The time-averaged value is conveniently estimated from the $\omega=0$ component of the spectrum. 

The different simulations we have carried out are documented in table \ref{tableconf3}.

\begin{table}[h]
\centering
\vspace{-2cm}
\begin{scriptsize}
\begin{tabular*}{\linewidth}{c @{\extracolsep{\fill}} lllll}	
	\toprule
		$\omega T$      & $N$ & $t$ & $\alpha_p$ &$R_\lambda$ \\
\colrule
		0       & 0  & 800         & 0.2   & 32\\    
		0.037 & 4    & 1316     & 0.2   & 32\\
		0.064 & 8    & 1519    & 0.2  & 32\\
		0.11  & 8    & 877    & 0.2  & 32\\
		0.19  & 8    & 506    & 0.2  & 32\\
		0.33  & 8    & 292    & 0.2  & 32 \\
		0.57  & 8    & 169    & 0.2  & 32\\
		1       & 15   & 183    & 0.2  & 32\\
		1.7   & 90   & 633    & 0.2  & 32\\
		3       & 180  & 731    & 0.2  & 32\\
		5  & 540  & 1266    & 0.2  & 32\\
		9       & 360  & 487    & 0.2  & 32\\
		16 & 810  & 633    & 0.2  & 32\\
		27      & 1080 & 487    & 0.2  & 32\\
		47 & 110  & 29     & 0.2   & 32\\
		81      & 200  & 30    & 0.2   & 32\\
		0       & 0    & 800         & 1   & 32\\    
		0.037 & 4    & 1316     & 1   & 32\\
		0.064 & 8    & 1519   & 1  & 32\\
		0.11  & 8    & 877    & 1  & 32\\
		0.19  & 8    & 506    & 1  & 32\\
		0.33  & 8    & 292    & 1  & 32 \\
		0.57  & 8    & 169    & 1  & 32\\
		1       & 15   & 183    & 1  & 32\\
		1.7   & 35   & 246    & 1  & 32\\
		3       & 60  & 243    & 1  & 32\\
		5  & 105  & 246    & 1  & 32\\
		9       & 180  & 243    & 1  & 32\\
		16 & 315  & 246    & 1  & 32\\
		27      & 540 & 243   & 1   & 32\\
		47 & 945  & 246     & 1   & 32\\
		81      & 1800  & 271     & 1   & 32\\
		0       & 0  & 40        & 1   & 105\\    
		0.11  & 2    & 262    & 1  & 105\\
		0.19  & 3    & 227    & 1  & 105\\
		0.33  & 3    & 131    & 1  & 105 \\
		0.57  & 3    & 76   & 1  & 105\\
		1       & 2   & 29   & 1  & 105\\
		1.7   & 6  & 50    & 1  & 105\\
		3       & 12  & 58    & 1  & 105\\
		5  & 54  & 151    & 1  & 105\\
		9       & 108  & 175    & 1  & 105\\
		16 & 54  & 50   & 1  & 105\\
		27      & 108 & 58    & 1   & 105\\
		47 & 162  & 50     & 1   & 105\\
		81      & 324  & 58    & 1   & 105\\
		\botrule
	\end{tabular*}
	\end{scriptsize}
	\caption{Simulation parameters: normalized frequency $\omega T$, number of simulated periods $N$, simulated time-interval $t$, relative forcing frequency $\alpha_p$ and Taylor-scale Reynolds number $R_\lambda$.  }
	\label{tableconf3}
\end{table}

\section{Response of turbulence and mixing on a periodic kinetic energy input \label{sec:flucU}} 

In this section we consider the case where we only modulate the kinetic energy,
\begin{eqnarray}
p=\overline{p} + \tilde p\cos(\omega t),\\
p_\theta =\overline{p_{\theta}}. 
\end{eqnarray}
It will be shown that the modulation $\tilde p$ of the velocity field does also affect the mixing of the passive scalar.

\subsection{Frequency response of the modulated kinetic energy and dissipation}

\begin{figure}[h]
	\begin{center}
	  \subfigure[]{\includegraphics[width=0.5\linewidth]{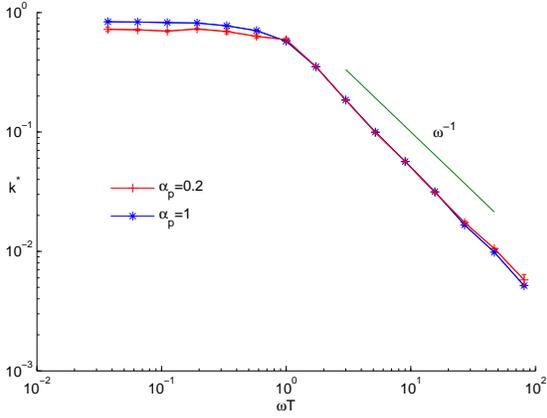}}~
	  \subfigure[]{\includegraphics[width=0.5\linewidth]{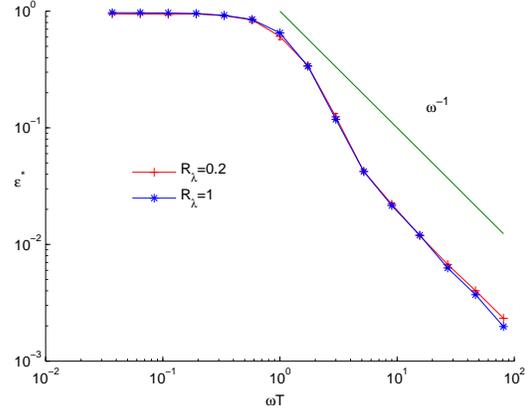}}
		\caption{The frequency dependence of (a) $k^*=\frac{\tilde k}{\overline k}\left(\frac{\tilde p}{\overline p}\right)^{-1}$ and (b) $\epsilon^*=\frac{\tilde \epsilon}{\overline \epsilon}\left(\frac{\tilde p}{\overline p}\right)^{-1}$ as a function of $\omega T$ for $R_{\lambda}=32$. Results for  $\alpha_p=0.2$ and $\alpha_p= 1$ are shown. \label{fig:normalized}}
	\end{center}
\end{figure}

The frequency response of $\tilde k$ and $\tilde \epsilon$ is shown in Fig. \ref{fig:normalized} for $R_\lambda=32$. We compare in this figure the frequency responses for two different relative forcing amplitudes, $\alpha_p=\tilde p/\bar p=0.2$ and $\alpha_p= 1$. In order to compare the frequency response for the different forcing amplitudes, we plot in these figures the quantities
\begin{equation}
 k^*=\alpha_p^{-1}\frac{\tilde k}{\overline k} \textrm{~~ and ~~} \epsilon^*=\alpha_p^{-1}\frac{\tilde \epsilon}{\overline \epsilon}
\end{equation}
as a function of frequency. Several observations can be made. Firstly, the $20\%$ and $100\%$ relative forcing amplitudes give results that superpose at almost all frequencies. In the following we will therefore focus on $\alpha_p=1$ results, which allow to obtain results at a lower computational cost, and therefore, at higher Reynolds number. Secondly, the powerlaw dependence proportional to $\omega^{-1}$ observed in \cite{Kuczaj2006} and \cite{Bos2007-3} is clearly reproduced both for $\tilde k$ and $\tilde \epsilon$. At small frequencies both $\tilde k$ and $\tilde \epsilon$ tend to constant values, as predicted in \cite{Bos2007-3}, but unlike the DNS results in \cite{Kuczaj2006} at these frequencies, perhaps due to unconverged statistics in their simulations. The quality of the collapse of the data at different values of $\alpha_p$ is not so good for $k^*$ in the low frequency limit. This is due to the value of $\overline k$ in this limit, which is affected by the strong modulation, as will be illustrated below, in section \ref{subsec:av}.

\begin{figure}[h]
	\begin{center}
	  \subfigure[]{\includegraphics[width=0.5\linewidth]{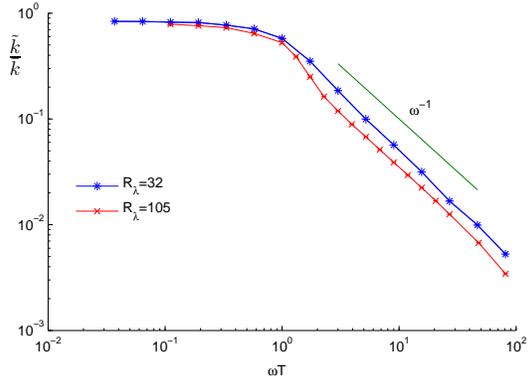}}~
	  \subfigure[]{\includegraphics[width=0.5\linewidth]{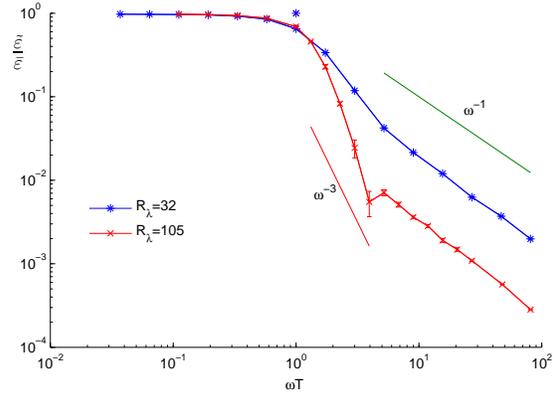}}	
		\caption{Frequency response of (a) the modulated kinetic energy ${\tilde k}/{\overline k}$  and (b) modulated dissipation  ${\tilde \epsilon}/{\overline \epsilon}$ for  $R_{\lambda}=32$($*$) and $R_{\lambda}=105$($\times$). The relative forcing amplitude is $\alpha_p= 1$.
		\label{fig:result1}}
	\end{center}
\end{figure}

Figure \ref{fig:result1} illustrates the influence of the Reynolds number on the modulated kinetic energy and dissipation. It is observed that this influence is small for the modulated kinetic energy. However, for the dissipation this influence is larger, as was explained in \cite{Bos2007-3} by the fact that the $\omega^{-1}$ asymptote is inversely proportional to the Reynolds number, since it corresponds to the direct influence of the viscous damping on the forced scales. The intermediate zone between the low frequency plateau and the high frequency asymptote is significantly steeper than the prediction that it should be proportional to $\omega^{-3}$ for large Reynolds numbers. Whether this is a low Reynolds number effect, or caused by the strong forcing, or due to something else can not be concluded from the present observations.

\begin{figure}
	\begin{center}
		  \subfigure[]{\includegraphics[width=0.5\linewidth]{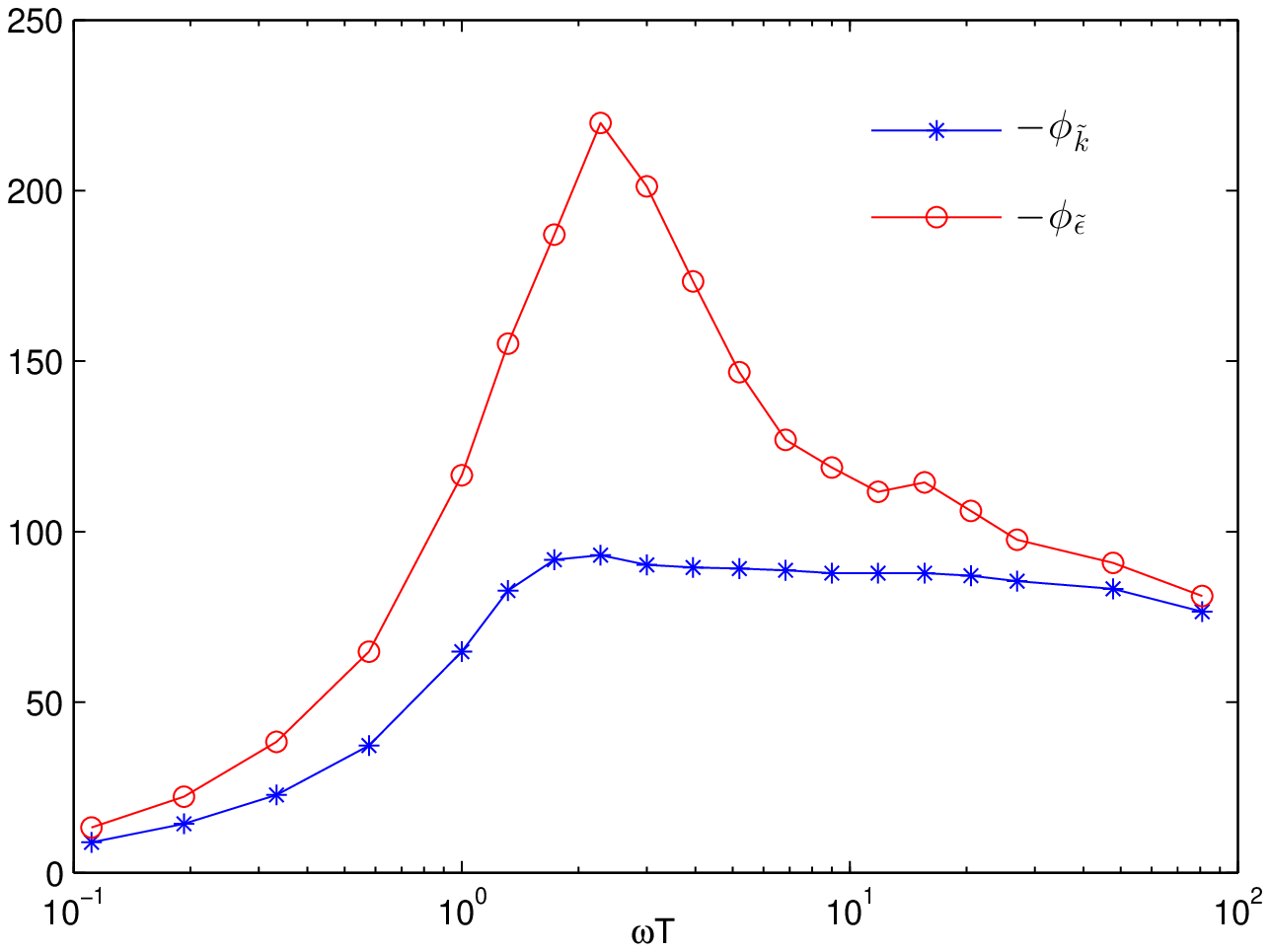}}~
		  \subfigure[]{\includegraphics[width=0.5\linewidth]{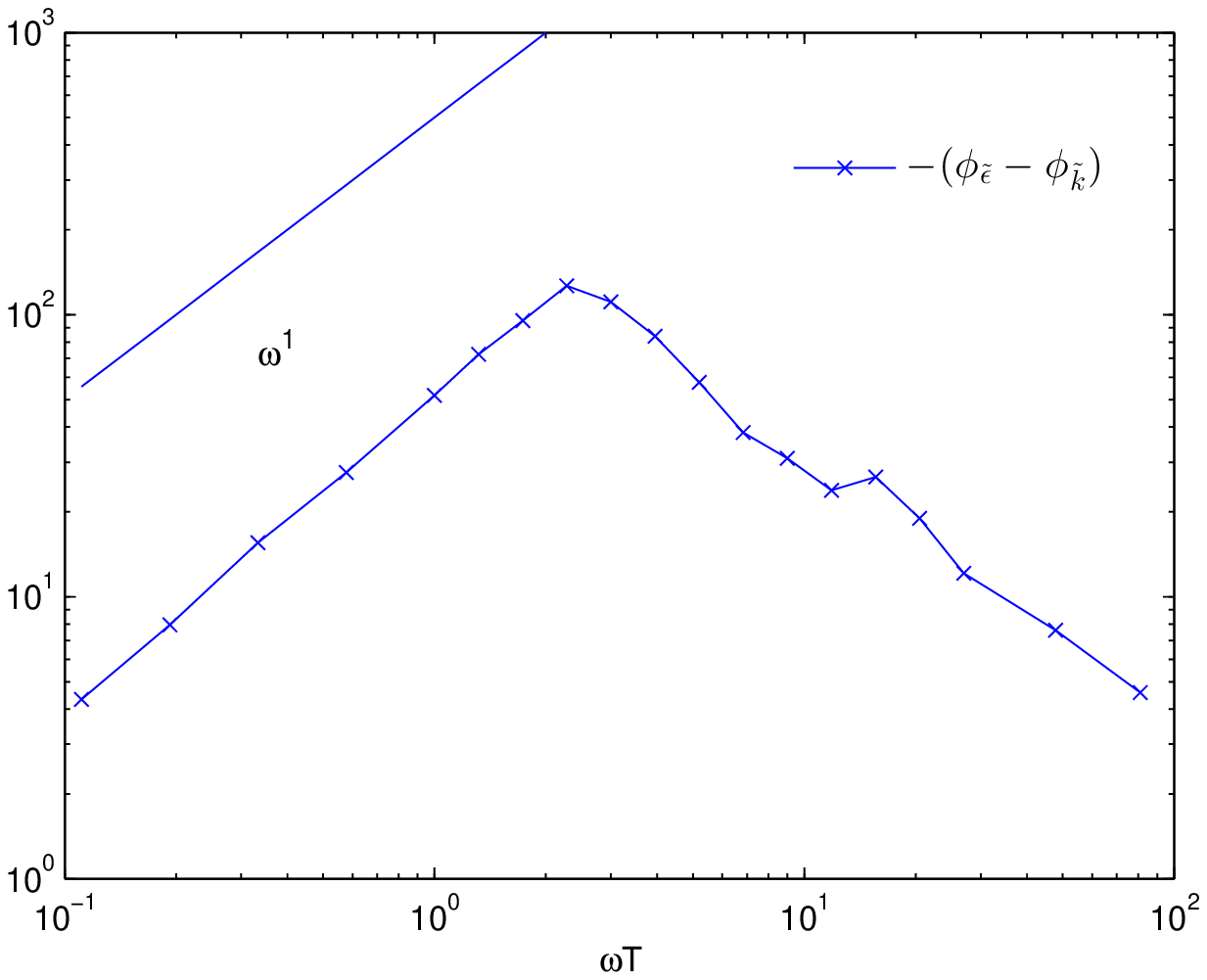}}	
	\end{center}
			\caption{(a) Phase shifts $ - {{\phi}_{\tilde k}}$ ($*$) and $ - {{\phi}_{\tilde {\epsilon}}}$ ($\circ$) as a function of $\omega T$ (in degrees) for $R_{\lambda}=105$ and $\alpha_p= 1$. (b) Relative Phase shift $ - ({{\phi}_{\tilde {\epsilon}}}- {{\phi}_{\tilde {k}}})$ as a function of $\omega T$ for $R_{\lambda}=105$ and $\alpha_p= 1$.}
			\label{fig:phase1and2}	
\end{figure}

In Figure \ref{fig:phase1and2} (a) the phase-shifts of the kinetic energy $-\phi_k$ and dissipation $-\phi_\epsilon$ with respect to the forcing are plotted. It is observed that $-\phi_k$ evolves almost continuously from $0$ at small frequencies, where everything is in phase, to the expected high frequency asymptote of $90\,^{\circ}$. A small overshoot is observed before reaching the high frequency asymptote, as also reproduced using closure simulations \cite{Bos2007-3} and in DNS \cite{Kuczaj2006}. The phase-shift $-\phi_\epsilon$ also behaves as expected, with a peak value around the integral frequency. The difference between the two phases is proportional to $\omega$ for small frequencies (Fig. \ref{fig:phase1and2} (b)), reflecting the finite cascade-time between the large and small scales of the flow \cite{Bos2007-3}. Interestingly a small local maximum is observed at $\omega T\approx 16$. We have currently no explanation for this observation and since is does not occur in the linear predictions, we suspect this maximum to be related to a violation of the linear-response assumptions.

Overall, these results show that the observations from previous studies are reproduced even though the relative forcing amplitude $\alpha_p$ is significantly higher in the present study.

\subsection{Frequency response of the time-averages \label{subsec:av}}

Within the framework of linear response theory, we consider that we measure the response of a system to an infinitesimal perturbation. In the case of a periodic perturbation the response is assumed to be at the same frequency as the perturbation, around the unaltered system. When the perturbation is large, the system itself can be affected importantly, and linear-response theory is no longer valid. In our case, a perturbation of the injection with an amplitude equal to the average injection cannot possibly be considered infinitesimal. It is therefore interesting to see if such a perturbation modifies the time-averaged properties of the flow.

\begin{figure}[h]
	\begin{center}
		  \subfigure[]{\includegraphics[width=0.5\linewidth]{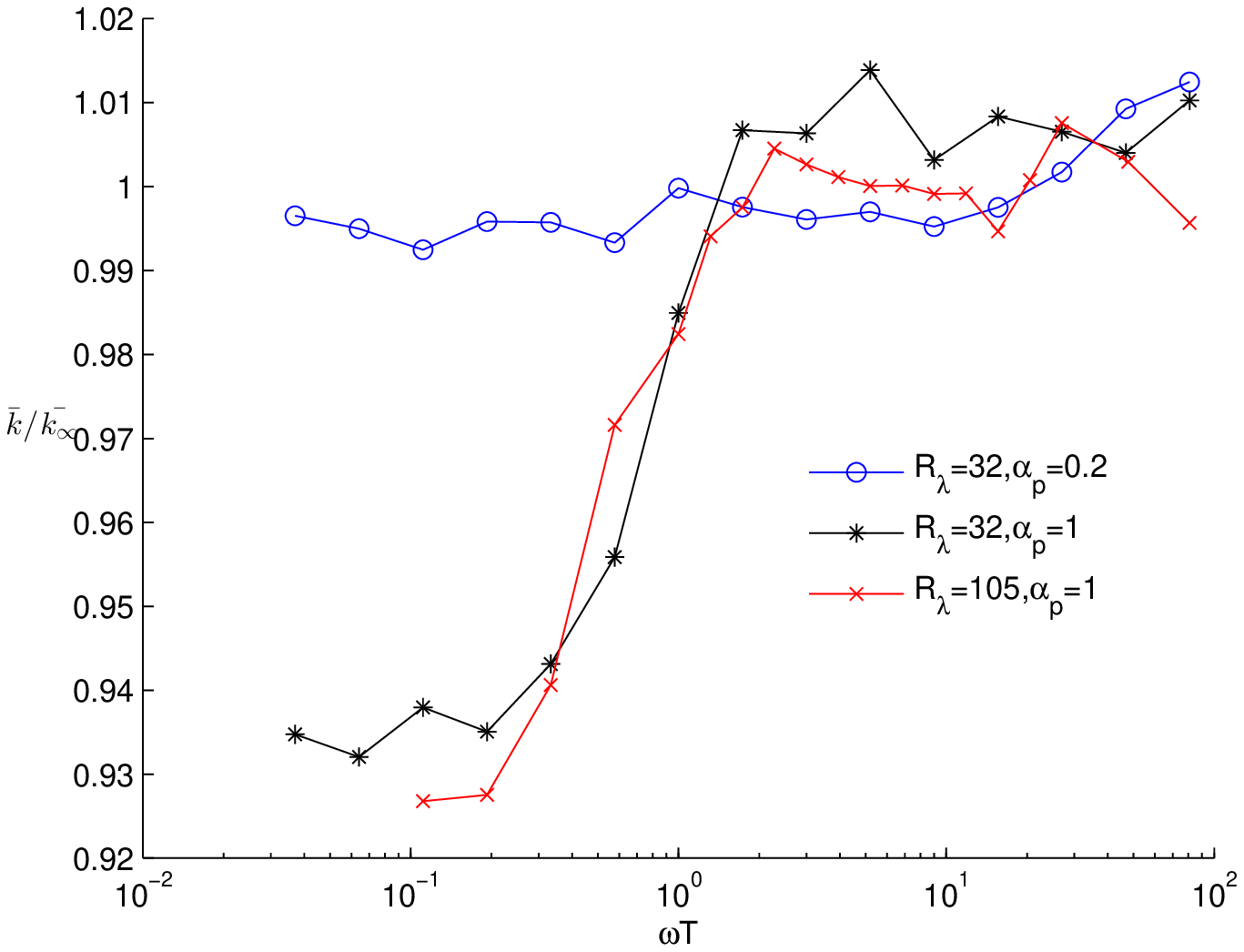}}~
		  \subfigure[]{\includegraphics[width=0.5\linewidth]{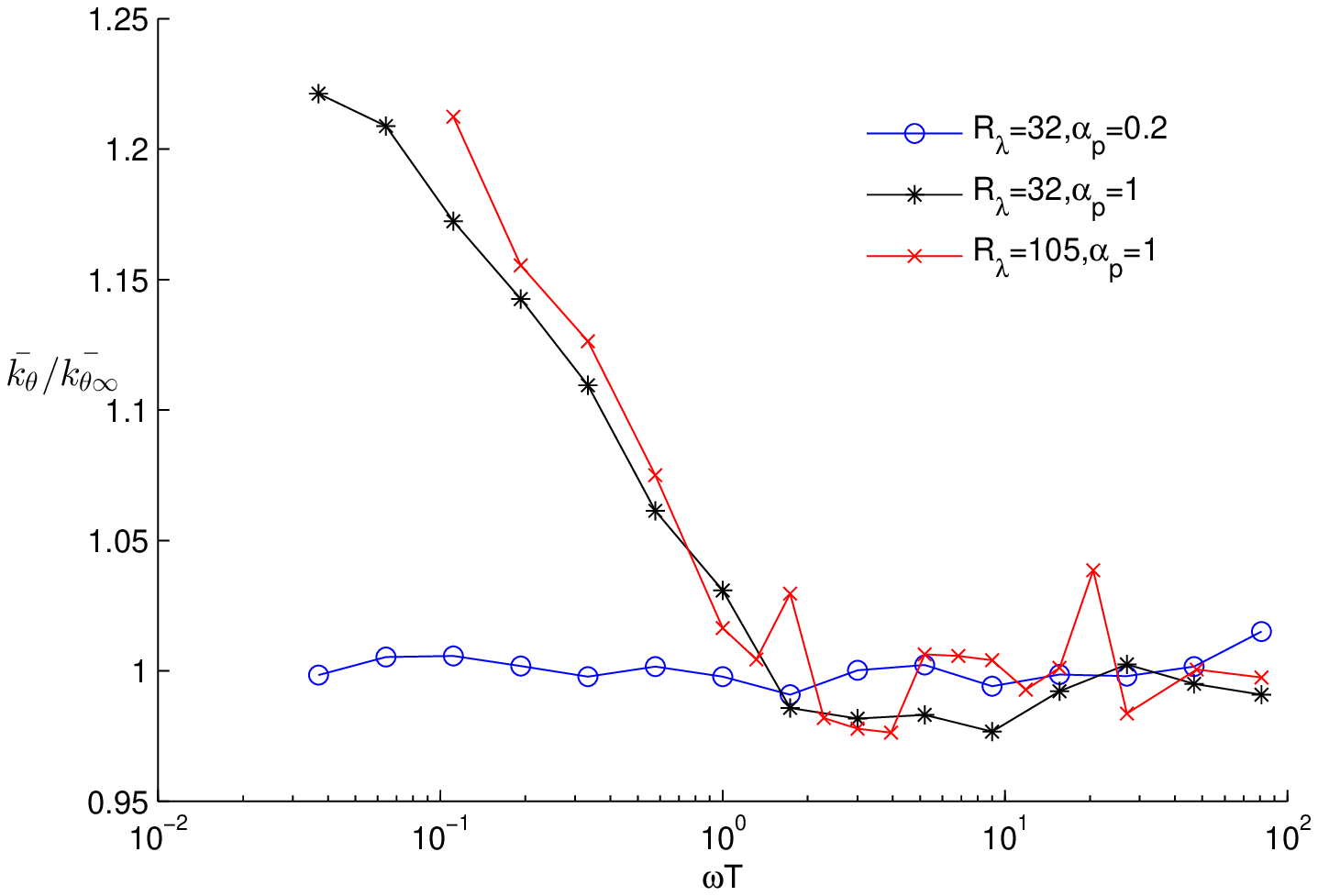}}
	\end{center}
		\caption{ Normalized time-averaged  kinetic energy $\bar{k}/\bar k(\omega=\infty)$ (a) and scalar variance $\bar{k_\theta}/\bar{k_\theta}(\omega=\infty)$ (b) as a function of $\omega T $. Results for  $R_{\lambda}=32$, $\alpha_p= 0.2$ ($\circ$); $R_{\lambda}=32$, $\alpha_p= 1$, ($*$); and $R_{\lambda}=105$, $\alpha_p= 1$  ($\times$).}
		\label{fig:constant2}
\end{figure}

We recall that in the present section $\tilde p_\theta=0$: only the forcing of the velocity contains a time-periodic contribution. Figure \ref{fig:constant2} shows the effect of modulated energy input on average components. It is observed that for low frequencies a quantitative deviation is observed from the average observed at high frequencies. This deviation is significantly larger than the error in the convergence of the statistics. It is in particular observed that $\overline k(\omega)/\overline k(\infty)$ decreases approximately to $93\%$ for small frequencies. On the contrary, the scalar variance increases more than $20\%$, i.e., $\overline k_\theta(\omega)/\overline k_\theta(\infty)\approx 1.2$. This effect is only observed for a forcing amplitude $\alpha_p=1$. For $\alpha_p=0.2$ the average seems unaffected, in agreement with the assumption of linear response. Also, two peaks in the frequency response of the average scalar variance are observed, one around the integral frequency, another one around $\omega T\approx 20$. The time-averaged dissipation $\overline \epsilon$ and scalar dissipation $\overline \epsilon_\theta$ (not shown) are not measurably affected. This is understandable since the forcing protocol is designed to maintain a well-defined value of these quantities.

Since the mixing rate is defined as $\chi_\theta=\bar \epsilon_\theta/\bar k_\theta$ (see equation (\ref{eq:mixeff})), the results in fig \ref{fig:constant2} show that the mixing-rate is negatively affected by the periodic forcing. But this effect disappears at high frequencies. The opposite is shown for the velocity field: the kinetic energy transfer rate is enhanced at low frequencies.  These results show that a periodic forcing can modify the mixing rate and the transfer rate when the modulation amplitude is large and the frequency is low. The requirement that the relative amplitude should be large to affect the averages is an indication that the effect is a nonlinear correction to the linear response prediction, where the time-average is not affected.

\subsection{Frequency response of the modulated scalar variance and dissipation}

\begin{figure}[h]
	\begin{center}
	  \subfigure[]{\includegraphics[width=0.5\linewidth]{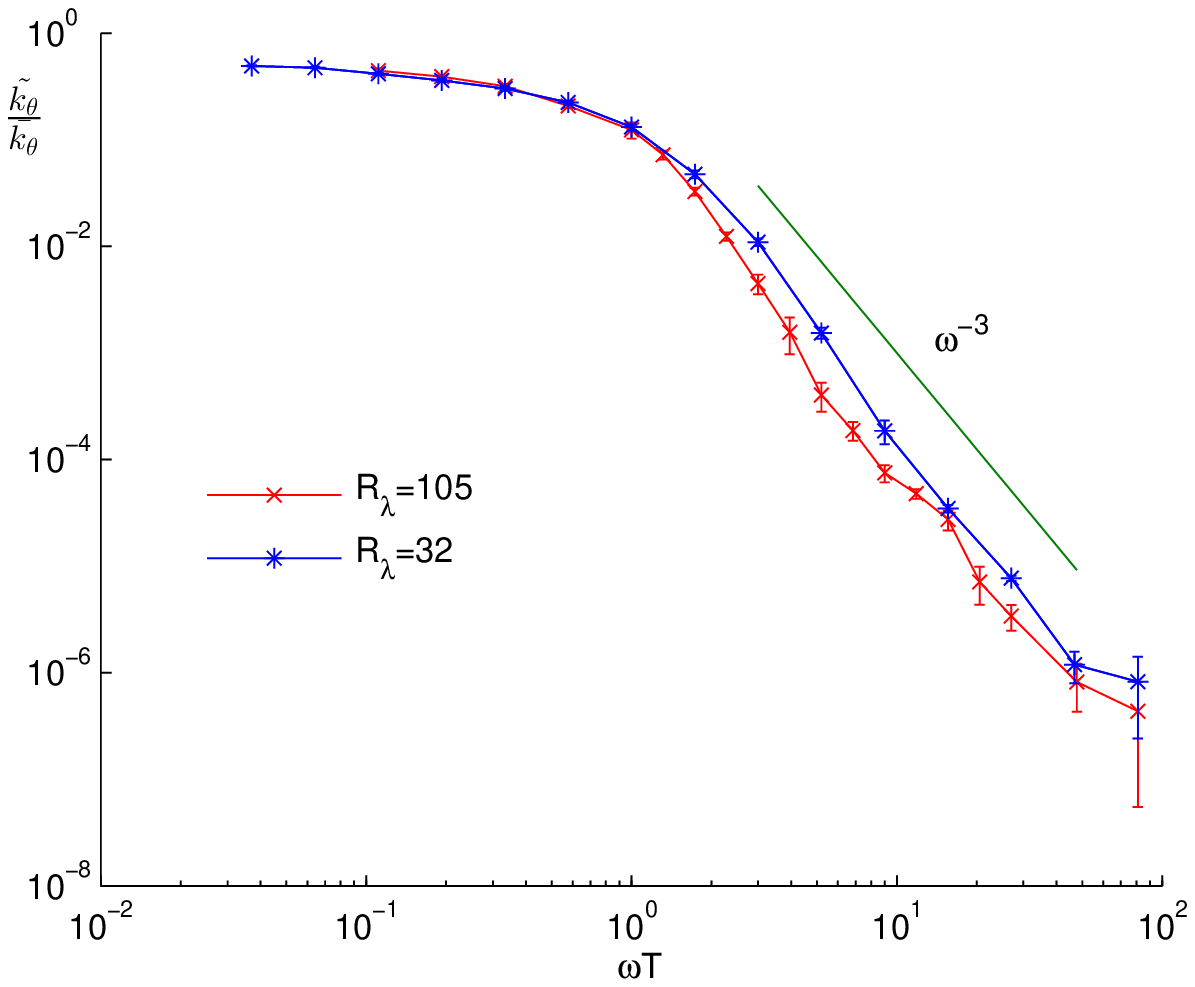}}~
	  \subfigure[]{\includegraphics[width=0.5\linewidth]{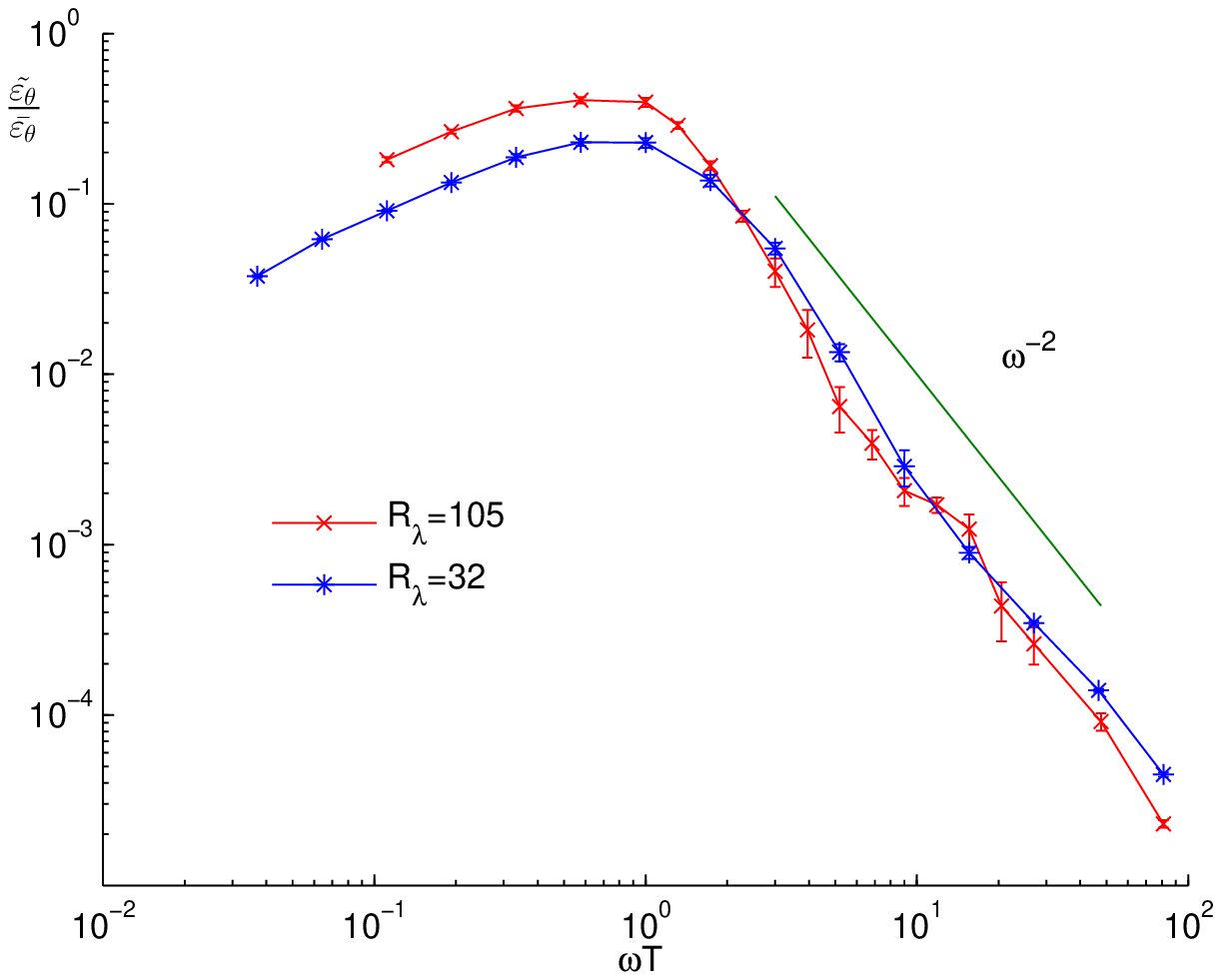}}
		\caption{Frequency response of (a)  the modulated scalar variance ${\tilde k_\theta}/{\overline k_\theta}$  and (b) modulated scalar dissipation ${\tilde \epsilon_\theta}/{\overline \epsilon}_\theta$ for  $R_{\lambda}=32$ ($*$) and $R_{\lambda}=105$ ($\times$). The relative forcing amplitude is $\alpha_p= 1$.
		\label{fig:result2}}
	\end{center}
\end{figure}

Since the scalar forcing does not contain a time-periodic component, the equation for the modulated scalar variance is,
\begin{eqnarray}
-\omega \tilde k_\theta \sin(\omega t+\phi_{k_\theta})=-\tilde \epsilon_\theta \cos(\omega t+\phi_{\epsilon_\theta}). \label{eq:ktheta0p}
\end{eqnarray}
From this equation it is not obvious that the scalar should contain a periodic component. Indeed from (\ref{eq:ktheta0p}), taking the modulus, we only find that
\begin{equation}
 \omega|\tilde k_\theta|=|\tilde \epsilon_\theta|.\label{eq:epsomegaktheta}
\end{equation}
No further information can be obtained from these equations about the frequency dependence of $\tilde k_\theta$. Obviously, $\tilde k_\theta=0$ is a possible solution of these equations. But since the velocity field  advecting the scalar contains a periodic contribution, it is not excluded that a periodic contribution is observed in the scalar dynamics. This is indeed the case, as observed in Figure \ref{fig:result2}.
It is observed that the scalar variance contains a periodic component which is constant at low frequencies, and rapidly drops off at high frequencies, following a powerlaw proportional to $\omega^{-3}$. The periodic part of the dissipation is then determined by relation (\ref{eq:epsomegaktheta}), as seems to be confirmed in figure \ref{fig:result2}. The origin of the $\omega^{-3}$ powerlaw is not easily obtained from the single point equations. 

\begin{figure}[h]
	\begin{center}
		  \subfigure[]{\includegraphics[width=0.5\linewidth]{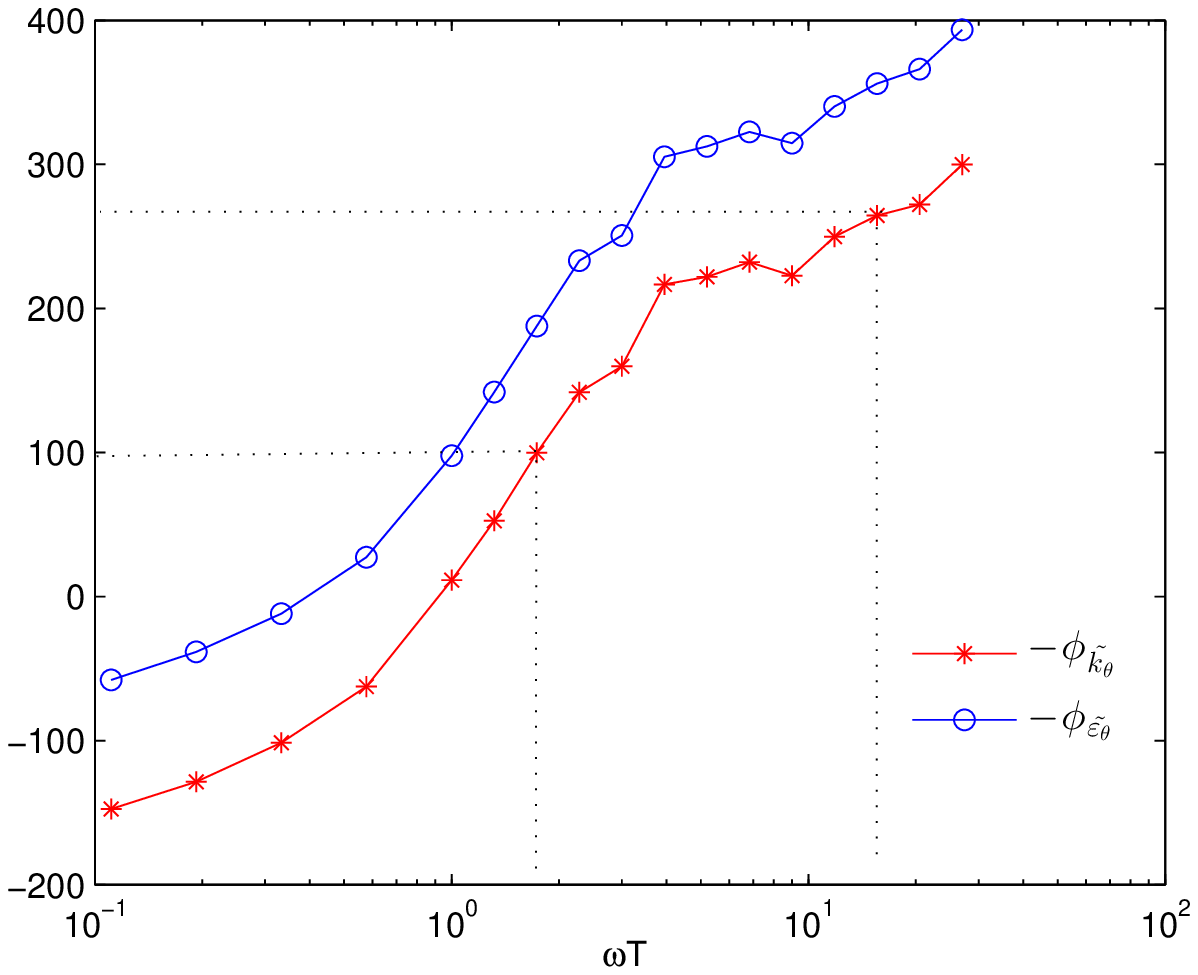}}~
		  \subfigure[]{\includegraphics[width=0.5\linewidth]{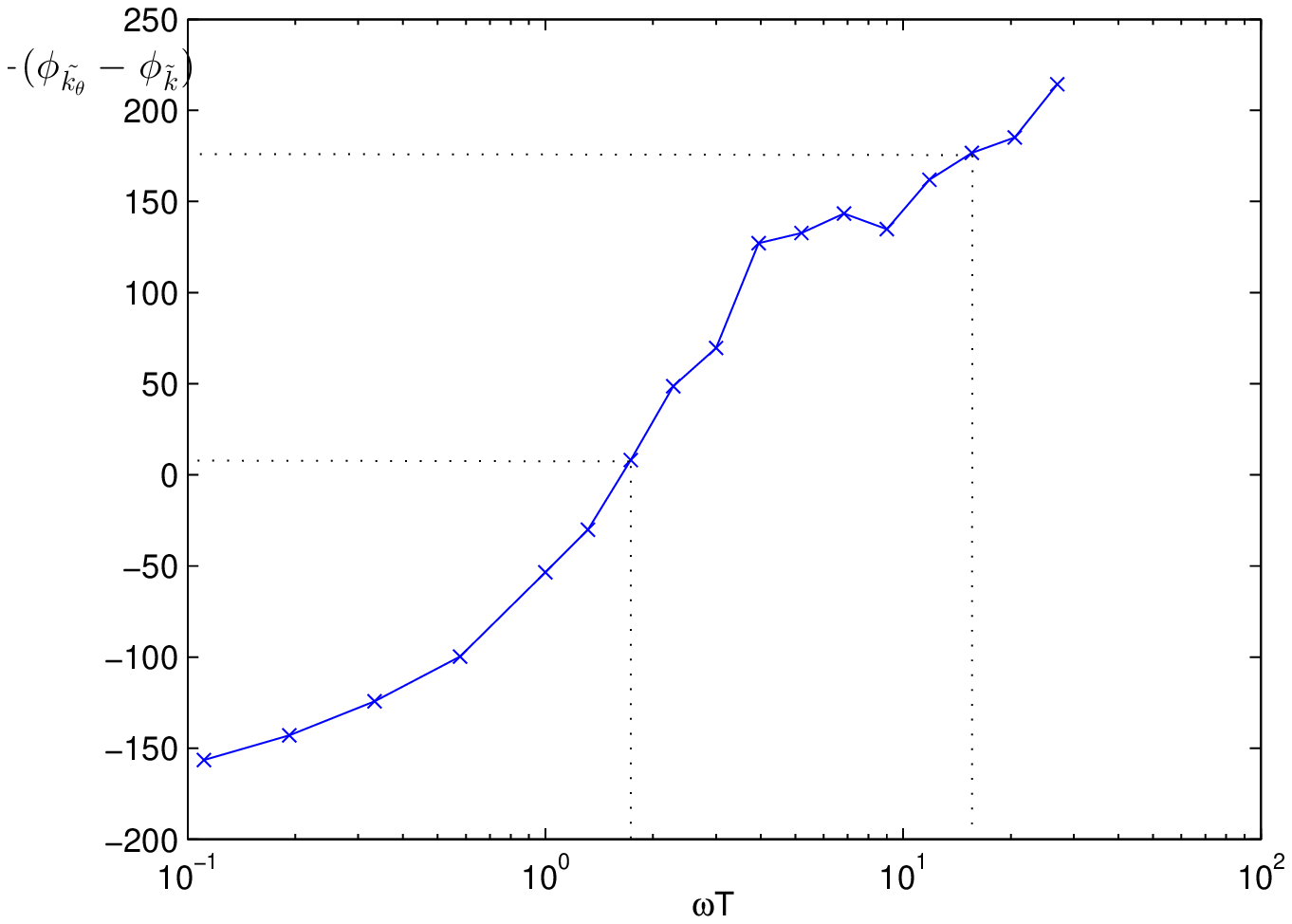}}
			\caption{(a) Phaseshifts $ - \phi_{k_\theta}$ ($*$) and $ - \phi_{\epsilon_\theta}$ ($\circ$) as a function of $\omega T$. (b) Phaseshift $ - (\phi_{k_\theta^2}- \phi_k)$. All results are given for $R_{\lambda}=105$ and $\alpha_p=1$.}
			\label{fig:phase3}	
	\end{center}
\end{figure}

In Figure \ref{fig:phase3} the phaseshifts $\phi_{\tilde k_\theta}$ and $\phi_{\tilde \epsilon_\theta}$ are shown. As expected from equation (\ref{eq:ktheta0p}) the relative phaseshift between $\tilde k_\theta$ and $\tilde \epsilon_\theta$ is always equal to $\pi/2$. We also consider the phaseshift between the signals of the modulated scalar variance and the modulated kinetic energy. If we evaluate this relative phaseshift at the two frequencies were the time-averaged scalar variance showed a clear peak (Fig. \ref{fig:constant2}) it is found that this corresponds to $ - ({\phi}_{k_\theta}- \phi_k)\approx 0$ or $\approx \pi$. It seems that a kind of resonance is observed in the scalar response, when the scalar variance and the kinetic energy are completely in phase, or completely out of phase.

\section{Modulation of the scalar injection \label{sec:flucT}}

\begin{figure}[h]
	\begin{center}
	  \subfigure[]{\includegraphics[width=0.5\linewidth]{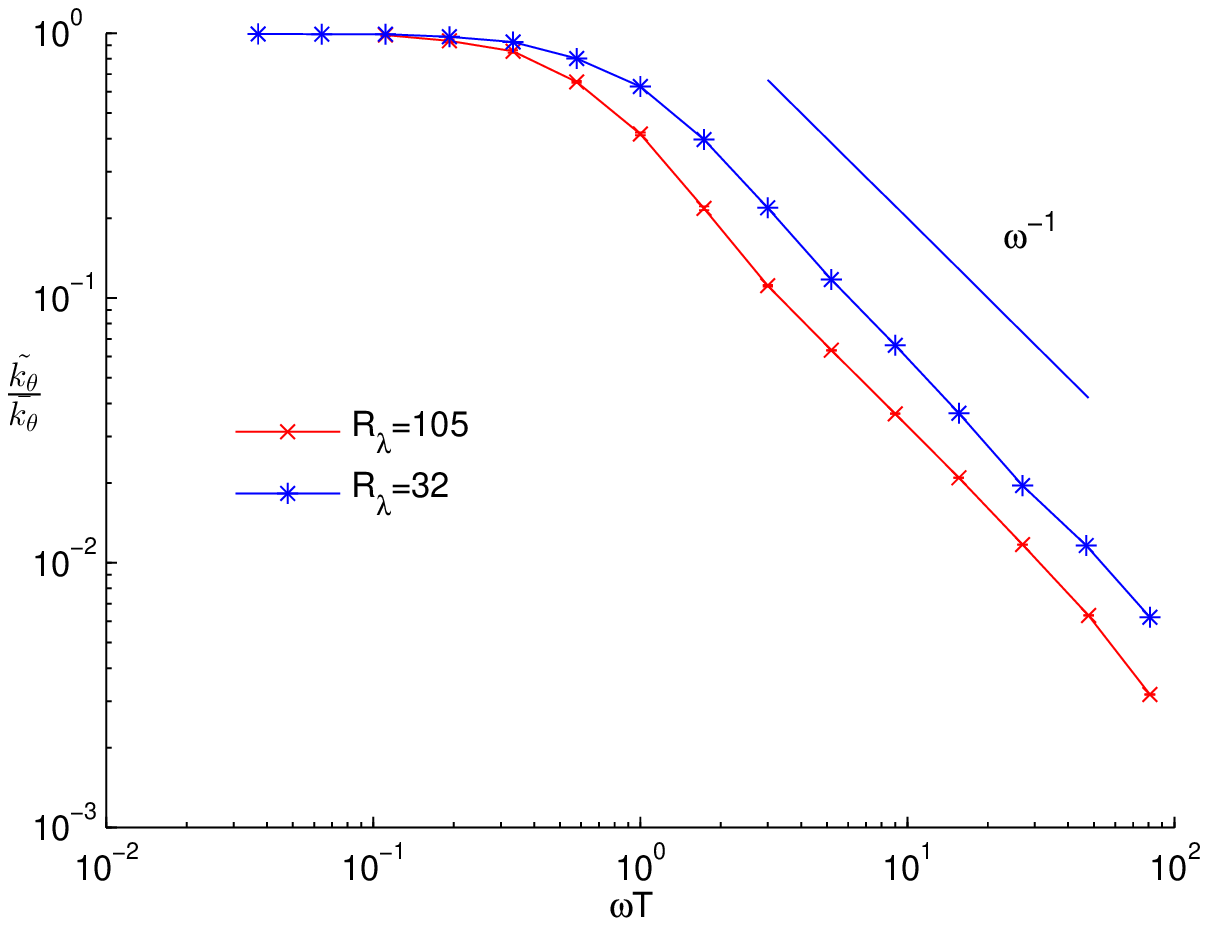}}~
	  \subfigure[]{\includegraphics[width=0.5\linewidth]{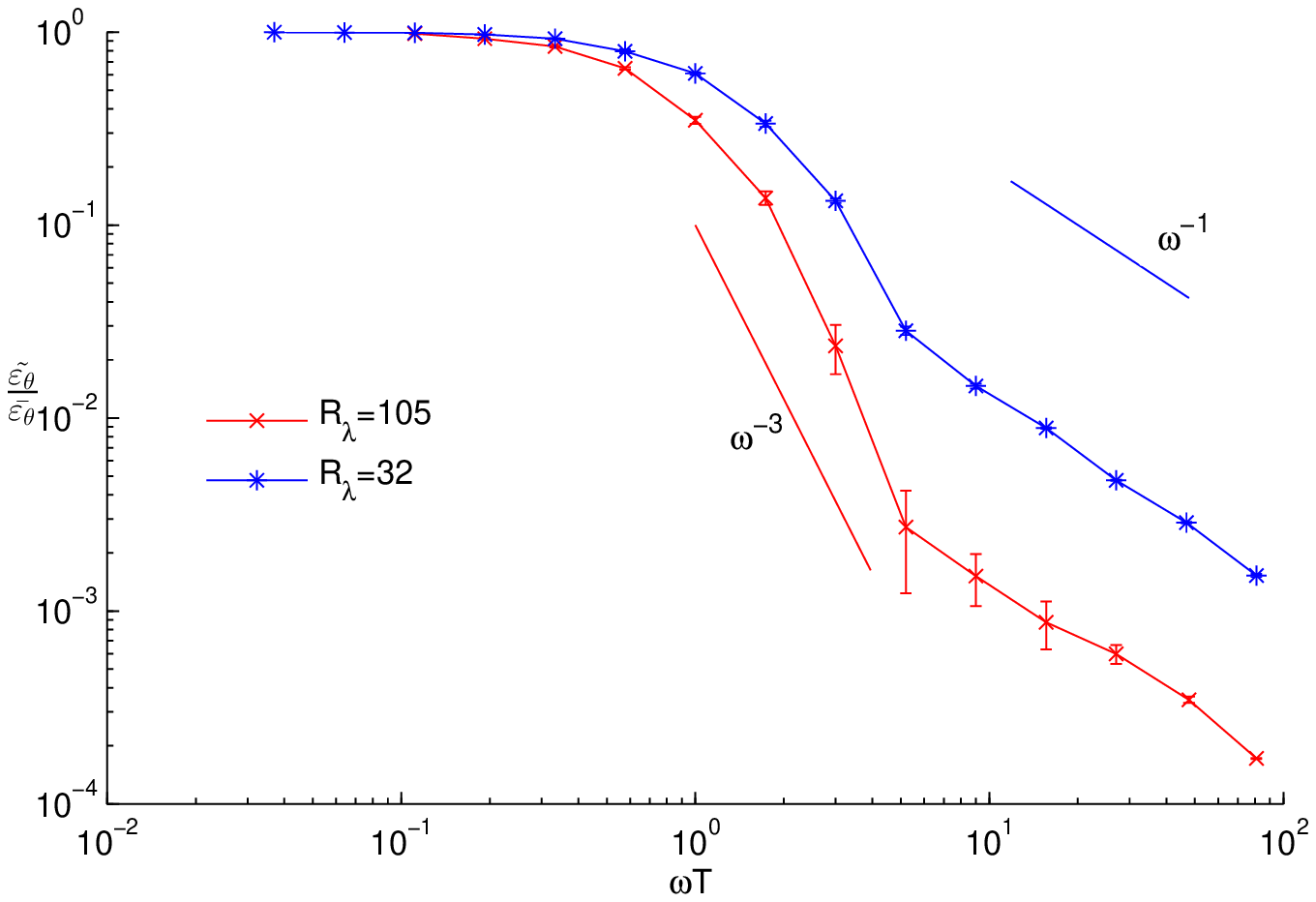}}
		\caption{ Amplitudes of the modulated scalar variance  $\tilde k_\theta/\bar k_\theta$ (a) and  dissipation $\tilde \epsilon_{\theta}/\bar\epsilon_\theta$ (b) as a function of $\omega T$ for the case of a modulated scalar injection. Results for $R_{\lambda}=32$($*$) and $R_{\lambda}=105$($\times$), both at $\alpha_p=1$.}
		\label{fig:result3}
	\end{center}
\end{figure}

In the foregoing, we have considered the case where the velocity was stirred at large scales by a periodic force term, but where the injection of the scalar was not modulated. In this section, we consider a modulated scalar injection in a non-time-periodic velocity field. In Figure \ref{fig:result3} we show the results for the modulated scalar variance $\tilde k_\theta/\bar k_\theta$ and dissipation $\tilde \epsilon_\theta/\bar \epsilon_\theta$. 

\begin{figure}	
	  \subfigure[]{\includegraphics[width=0.5\linewidth]{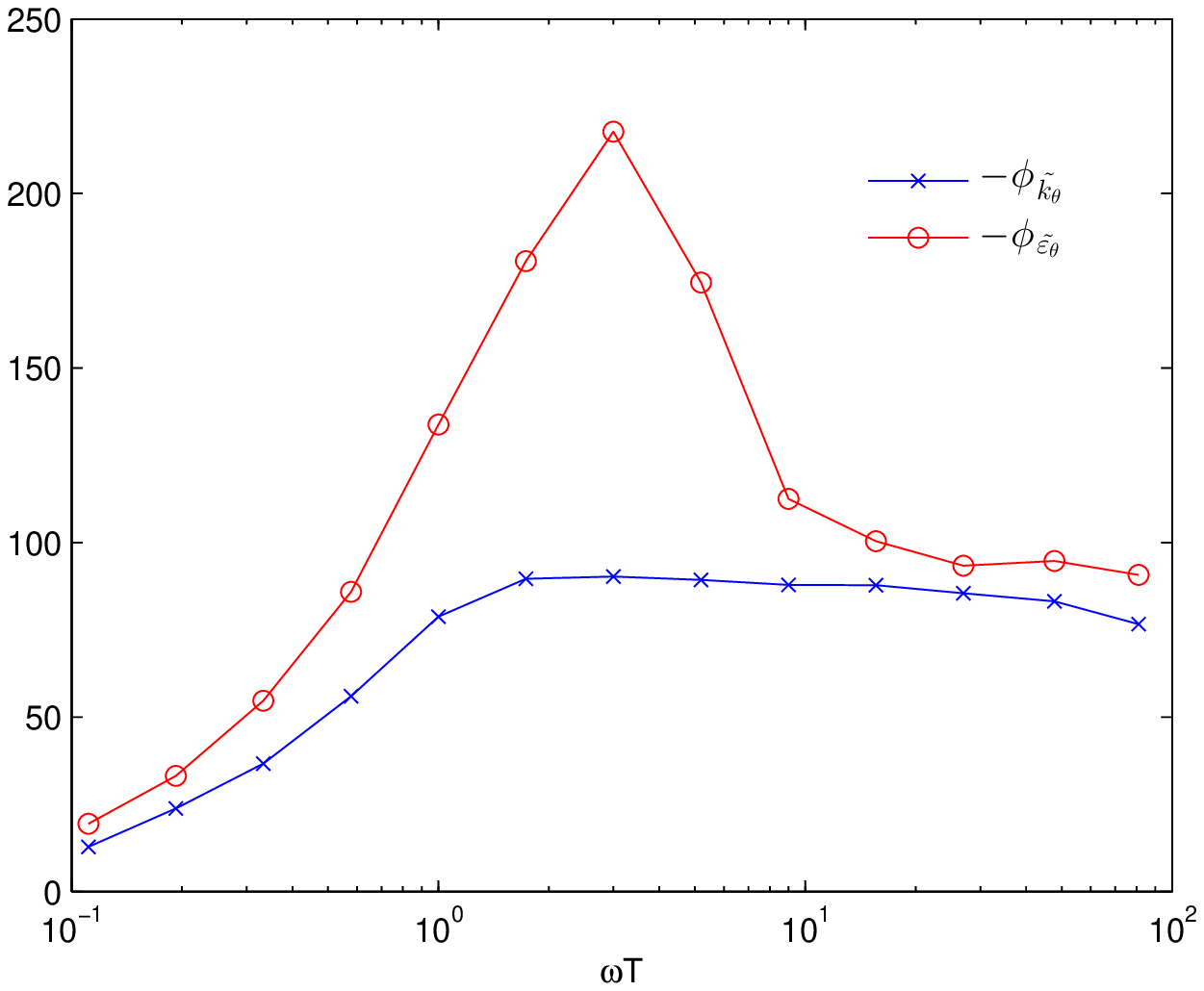}}~
	  \subfigure[]{\includegraphics[width=0.5\linewidth]{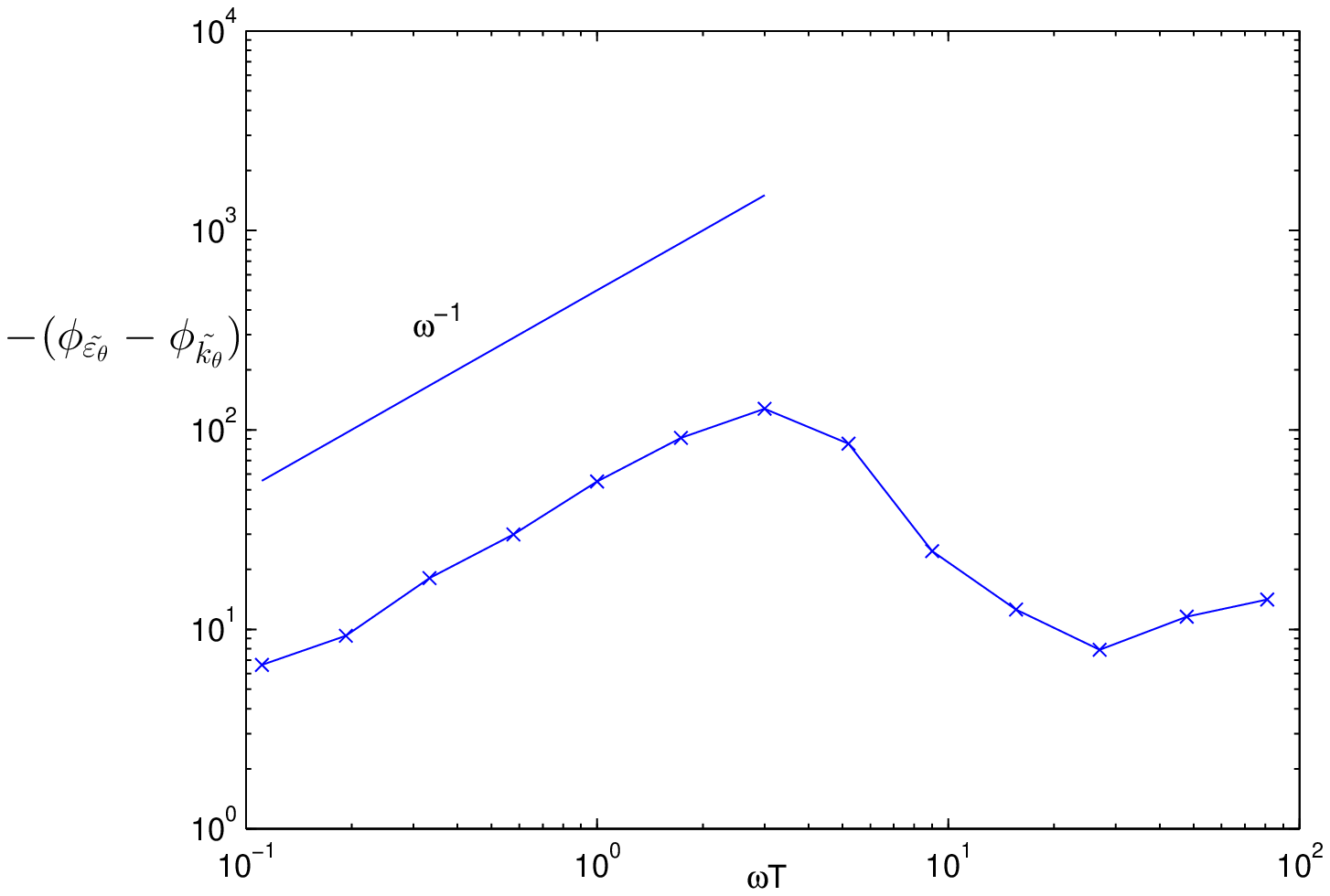}}
			\caption{(a) Phase shifts $ - \phi_{k_\theta}$ ($*$) and $ - \phi_{\epsilon_\theta}$ ($\circ$) as a function of $\omega T$. (b) Relative Phase shift $ - (\phi_{\epsilon_\theta}- \phi_{k_\theta})$. Results for $R_{\lambda}=105$ and $\alpha_p= 1$.}
			\label{fig:phase5}	
\end{figure}

A clear resemblance with the frequency behaviour of the modulated kinetic energy and dissipation, in Figure \ref{fig:result1}, is observed.  In particular the small and large frequency asymptotes are identical. Indeed, the reasonings leading to the prediction of the frequency behaviour of the kinetic energy and dissipation \cite{Bos2007-3} can be extended to the case of the passive scalar, in particular in the limit of the linear response approximation. A further resemblance is observed in the phase-shift of the scalar variance and dissipation, shown in figure  \ref{fig:phase5}. It seems that the results of the periodically forced velocity can be transposed to the periodically forced scalar, without major modifications.

\section{Conclusion}

Perhaps the most interesting outcome of the present investigation is that the  periodical modulation of a turbulent velocity field can directly affect its mixing rate. This effect was only observed when the relative forcing frequency was large ($\alpha_p= 1$), suggesting that the effect is nonlinear.  The average mixing rate was shown to decrease by approximately 20\%, for two distinct Reynolds numbers, whereas the energy transfer rate increases by approximately 10\%. Indeed, in the linear limit this effect is absent. 

We further showed that in this nonlinear regime the frequency response of the amplitudes of the kinetic energy and dissipation were still in agreement with the outcome from linear perturbation analysis. Interestingly, the modulated velocity also directly induces a modulation of the scalar field. It is therefore, according to our investigation possible to modify the time-averaged and phase-averaged scalar by acting on the velocity field.  

By adding a modulation on the scalar injection rate, it was shown that the frequency response of the periodic contributions of the scalar variance and the scalar dissipation behave very similar to the kinetic energy and dissipation for the case of the velocity modulation.


\end{document}